\documentclass[12pt,twoside]{article}
\usepackage{fleqn,espcrc1}
\usepackage{epsfig,graphicx}

\newcommand{\AmS}{{\protect\the\textfont2
  A\kern-.1667em\lower.5ex\hbox{M}\kern-.125emS}}

\title{
Relativistic NN scattering equations without 
partial-wave decomposition}

\author{G. Ramalho$^a$, 
        A. Arriaga\address{Centro de F{\'\i}sica Nuclear da 
Universidade de Lisboa, 1649-003 Lisboa Codex, Portugal.}
\address{Faculdade de Ci\^encias da Universidade de Lisboa, 1749-016 Lisboa, 
Portugal.}
 and  
M. T. Pe\~na\address{CFIF, 
Centro de F{\'\i}sica das Interac\c c\~oes Fundamentais,
IST, 1049-001 Lisboa, Portugal.}
\address{Dep.\ de F{\'\i}sica, IST, 
1049-001 Lisboa, Portugal.}
}

\begin{document}

\maketitle

\begin{abstract}
Relativistic quasi-potential equations 
describing NN scattering are compared. 
Within the spectator 
formalism a cancellation is seen to occur
between retardation and 
negative-energy effects. 
\end{abstract}

\section{INTRODUCTION}

Modern experimental facilities such as TJNAF and Julich
are probing the baryonic structure and interactions at 
intermediate energies ($\sim $ GeV), through electromagnetic and strong
reactions involving high momentum 
transfer \cite{Burkert00}.
To describe the NN interaction in the 
intermediate energy range ($\sim $ 300 MeV-1 GeV),  
particle production mechanisms have to be included.  
Relativistic effects became also important.
In this scenario an accurate relativistic  
treatment of NN scattering, including boosts and retardation effects,
is needed.
These dynamical aspects are included automatically when one solves de 4-dimensional 
Bethe-Salpeter 
integral equation \cite{Bethe-Salpeter}:
\begin{equation}
T(p^\prime,p;P)=V(p^\prime,p;P)
+i \int \frac{d^4 k}{(2\pi)^4} 
V(p^\prime,k;P) g(k;P) T(k,p;P).
\label{scatteringEq}
\end{equation}
%----------------------------------------- Figure --------------------
\begin{figure}[h]
\centerline{
\vspace{-0.5cm}
\epsfig{file=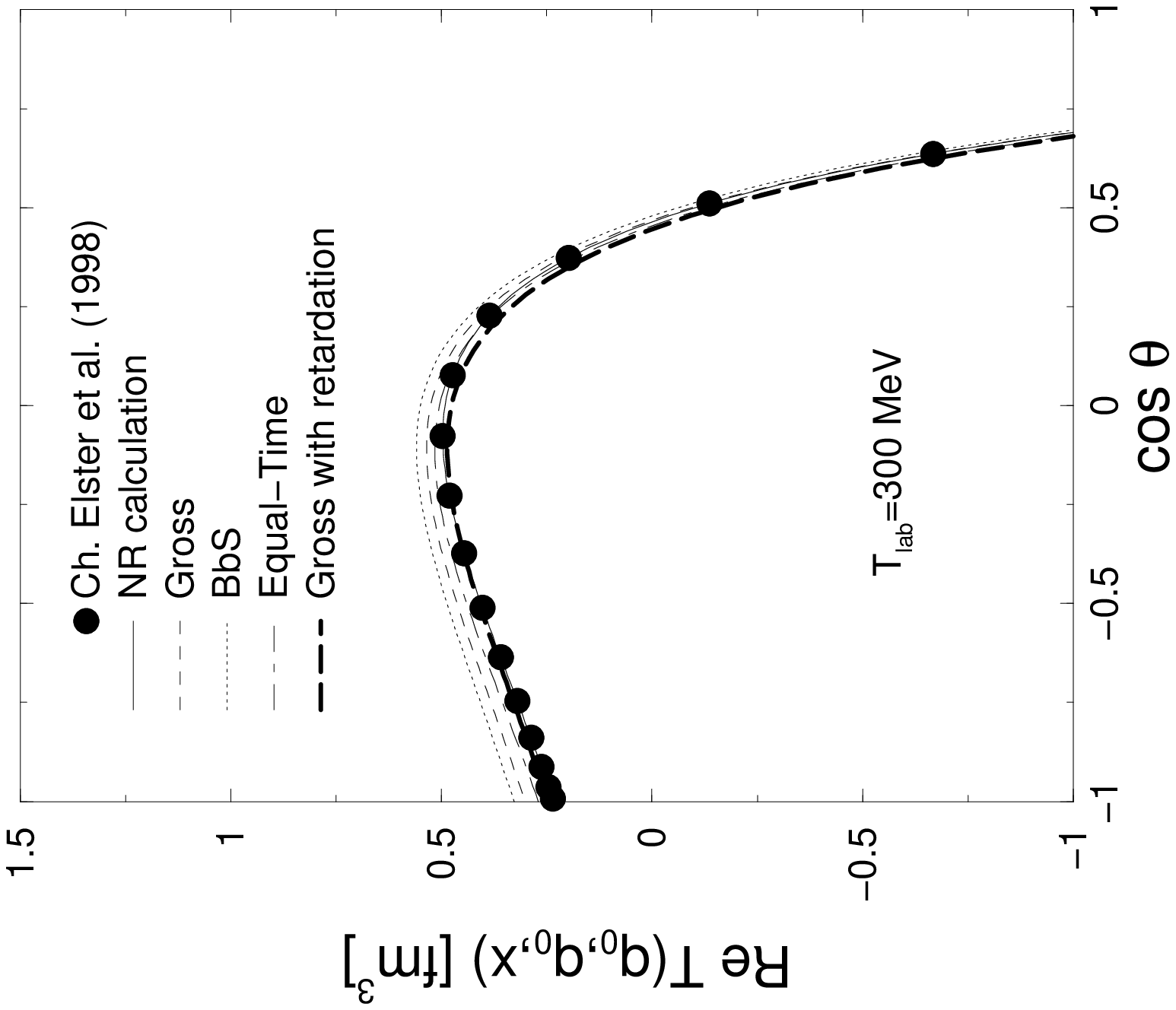, width=7.5cm, angle=270 } 
\epsfig{file=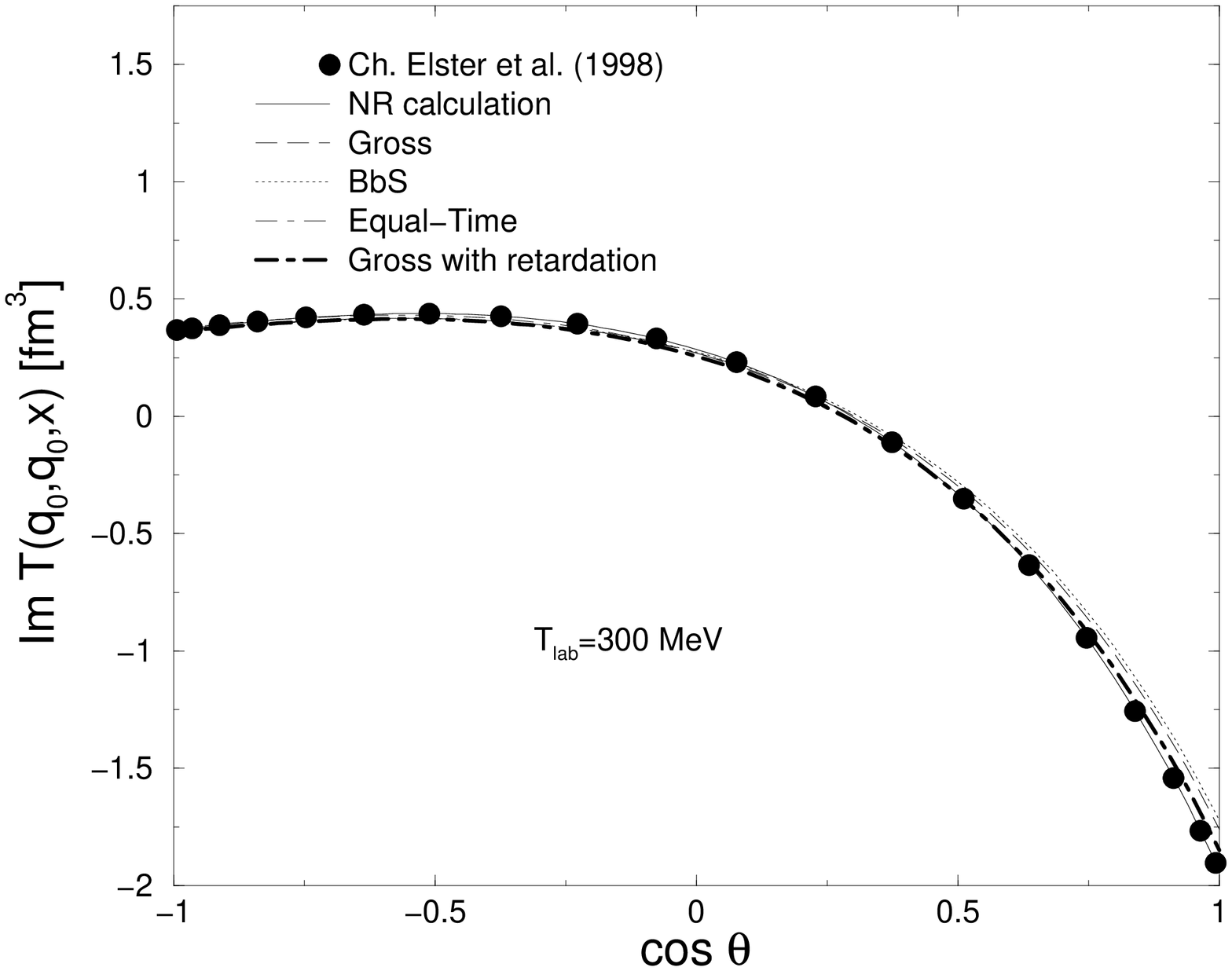, width=7.5cm, angle=270}} 
\vspace{-0.5cm}
\caption{300 MeV scattering amplitude. 
The bullets are from Ref.\ [6]. The solid line 
corresponds to a calculation 
with more mesh points.}
\end{figure}

We tested different 3-dimensional reductions of this equation, 
currently known as Quasi-Potential (QP) equations. 
These equations  
corresponds to different choices of the relative interaction energy. 
The QP equations considered are: the Gross or spectator equation [3]
(one particle on-mass-shell in all intermediate states), the 
Blankenbecler-Sugar equation (BbS) [4] 
(both the two particles equally off-mass-shell in all intermediate states,
which means 
no retardation effects), 
and the Equal-Time equation [5] (BbS with effects from
crossed-box diagrams included partially). 
The corresponding propagators for the intermediate states are 
respectively:
\begin{eqnarray}
g_{BbS}(k;P)&=& i 2 \pi \frac{M}{E_k}
\frac{M \delta  \left(k_0\right)}
{E_k^2-\frac{W^2}{4}-i\varepsilon}, \\
g_{ET}(k;P)&=&i 2 \pi \frac{M}{E_k}
\frac{M \delta  \left(k_0\right)}
{E_k^2-\frac{W^2}{4} -i \varepsilon} \times 
\left(2-\frac{W^2}{4 E_k^2} \right), \\
g_{Gross}(k;P)&=& i 2 \pi \frac{M}{E_k} \frac{M}{W}
\frac{M \delta  \left(k_0 +W/2-E_k\right)}
{E_k-\frac{W}{2}-i \varepsilon}, 
\end{eqnarray}
where $m$ is the 
nucleon mass, $E_k=\sqrt{m^2+{\bf k}^2}$ and $W=\sqrt{P^2}$.

\begin{figure}[t]
\centerline{
\vspace{-0.5cm}
\epsfig{file=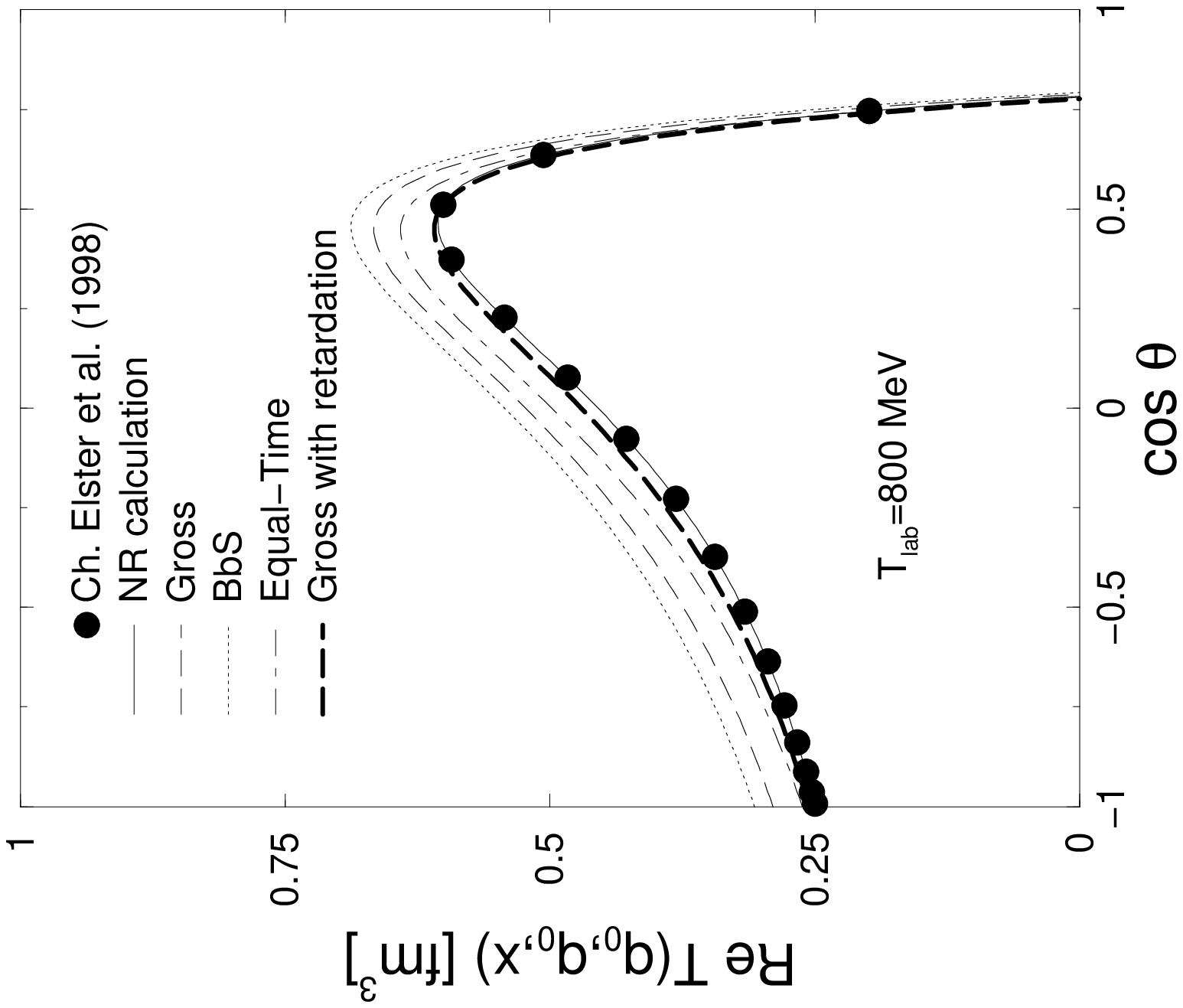, width=7.5cm, angle=270} 
\epsfig{file=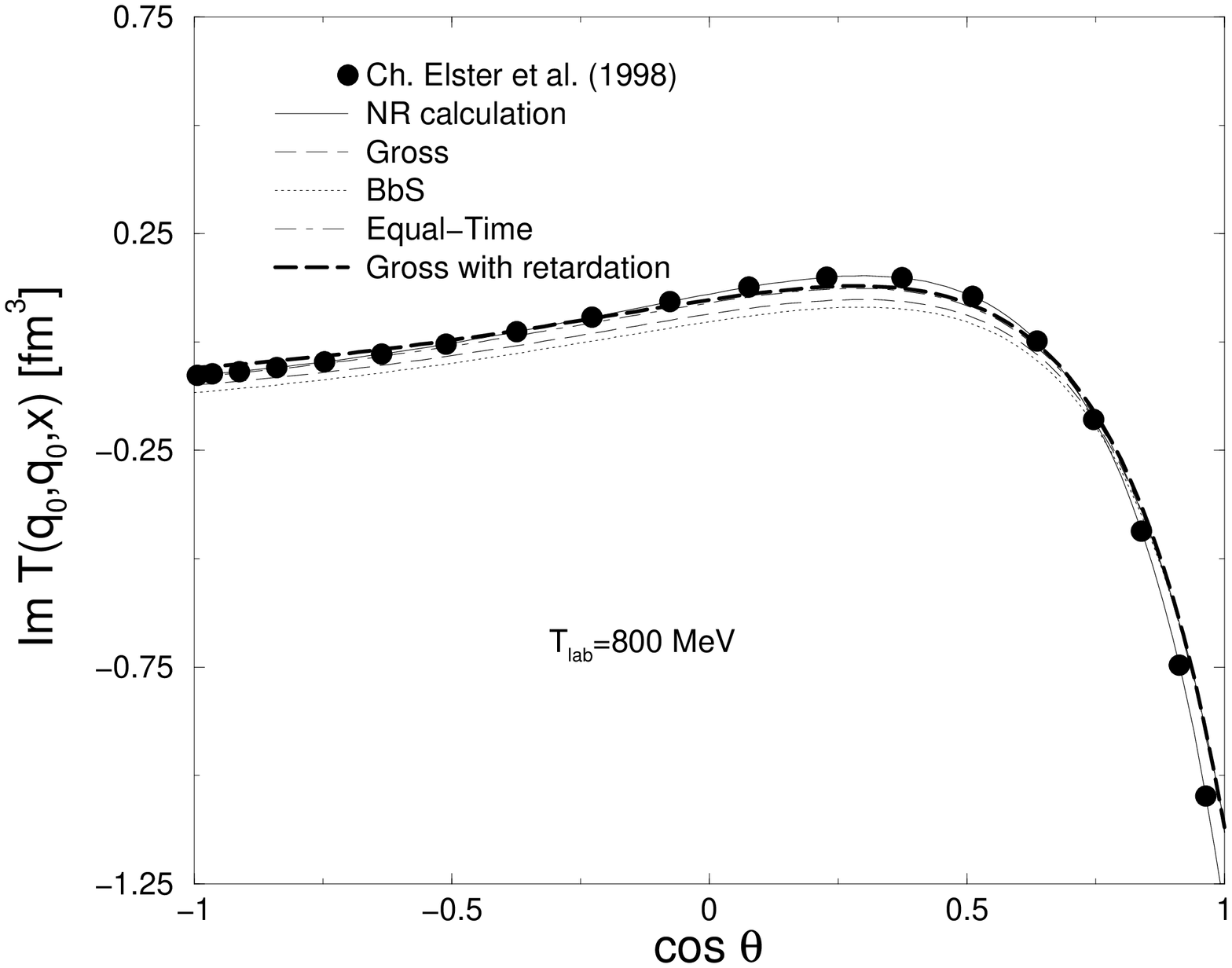, width=7.5cm, angle=270}} 
\vspace{-0.5cm}
\caption{800 MeV scattering amplitude. 
The bullets are from of Ref.\ [6]. The solid line 
corresponds to a calculation with 
more mesh points.}
\end{figure}

\section{PARTIAL WAVE-DECOMPOSITION}

Usually the 3-dimensional equations are solved by implementing 
a partial-wave decomposition in the angular variables. 
However, from the calculational point of view, the 
partial-wave decomposition of the NN scattering amplitude
becomes less adequate and practical for high energies [6], 
where an increasing number of partial-waves is required. 
Fortunately, nowadays computational resources allow the
evaluation of the scattering amplitude in terms of the on-shell  
momentum and scattering angle, without performing
a partial-wave expansion. This one is traded for the resolution of 
an integral equation in two dimensions, instead of only one, after
factoring out the integral
in the azimuthal variable $\varphi$. 
Without using a decomposition in partial-waves,  
Ch.\ Elster et al.\ \cite{Elster98} solved already the Non Relativistic (NR) 
Lippman-Schwinger 
scattering equation for scalar particles interacting through the 
Malfliet-Tjon potential. 
Subsequently, they used the two-nucleon half-off-shell amplitudes in 
a three-nucleon system calculation [7]. 
We have reproduced their NR two-nucleon results. 
For that purpose we used non-relativistic relations 
between laboratory and the CM frames. 
Moreover, we compared them with results obtained from     
different relativistic scattering equations.  
We also compared the NR results with 
the ones obtained from de three different 
relativistic equations mentioned above.
The results are presented in Figs.\ 1 and 2.

\section{RESULTS}

Our results quantify the increasing importance of 
relativistic effects with increasing 
energy. We have considered two energies cases: one (300 MeV)
in the {\it low} energy and another (800 MeV) in the {\it high} energy region.  
The relativistic effects turn to be more significant for the BbS 
and spectator/Gross equations, as we can see in Figs.\  1 and 2, 
particularly in the real part of the scattering amplitude: indeed,
the Equal-Time amplitudes are the ones
which are closer to the NR Lippman-Schwinger amplitudes. 
Deviations from the NR results increase with the energy. 

As for retardation effects, we modified the Malfliet-Tjon 
such that the scalar exchanges would consider the transfered energy. 
By construction these effects can be present only 
in the spectator/Gross QP equation.
With retardation effects included, 
the spectator  
amplitude (thick dashed line in Figs.\ 1 and 2) becomes closer to the 
NR limit than the Equal-Time amplitude.

%----------------------------------------- Figure --------------------
\begin{figure}[t]
\centerline{
\epsfig{file=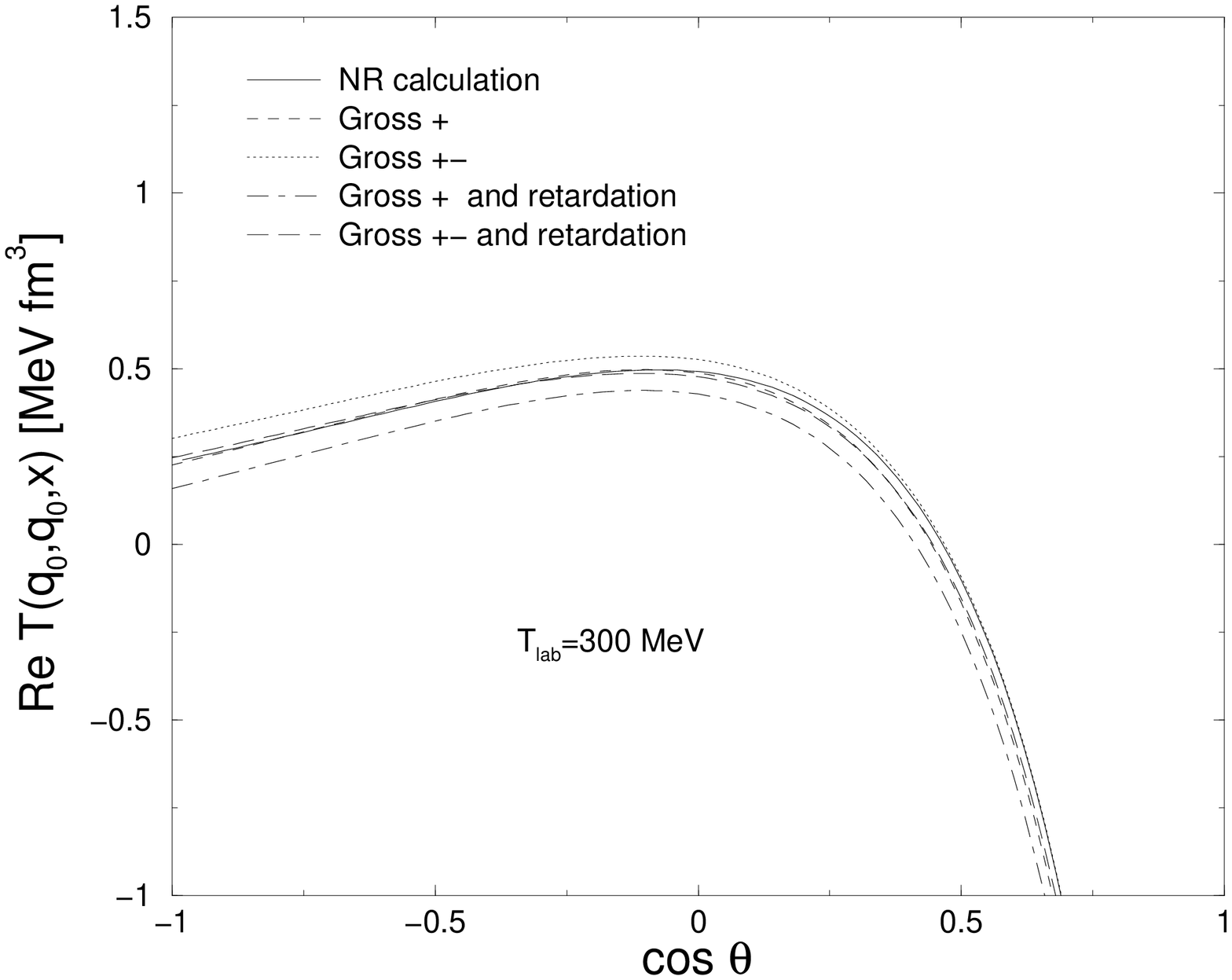, width=6.5cm, angle=270} 
\epsfig{file=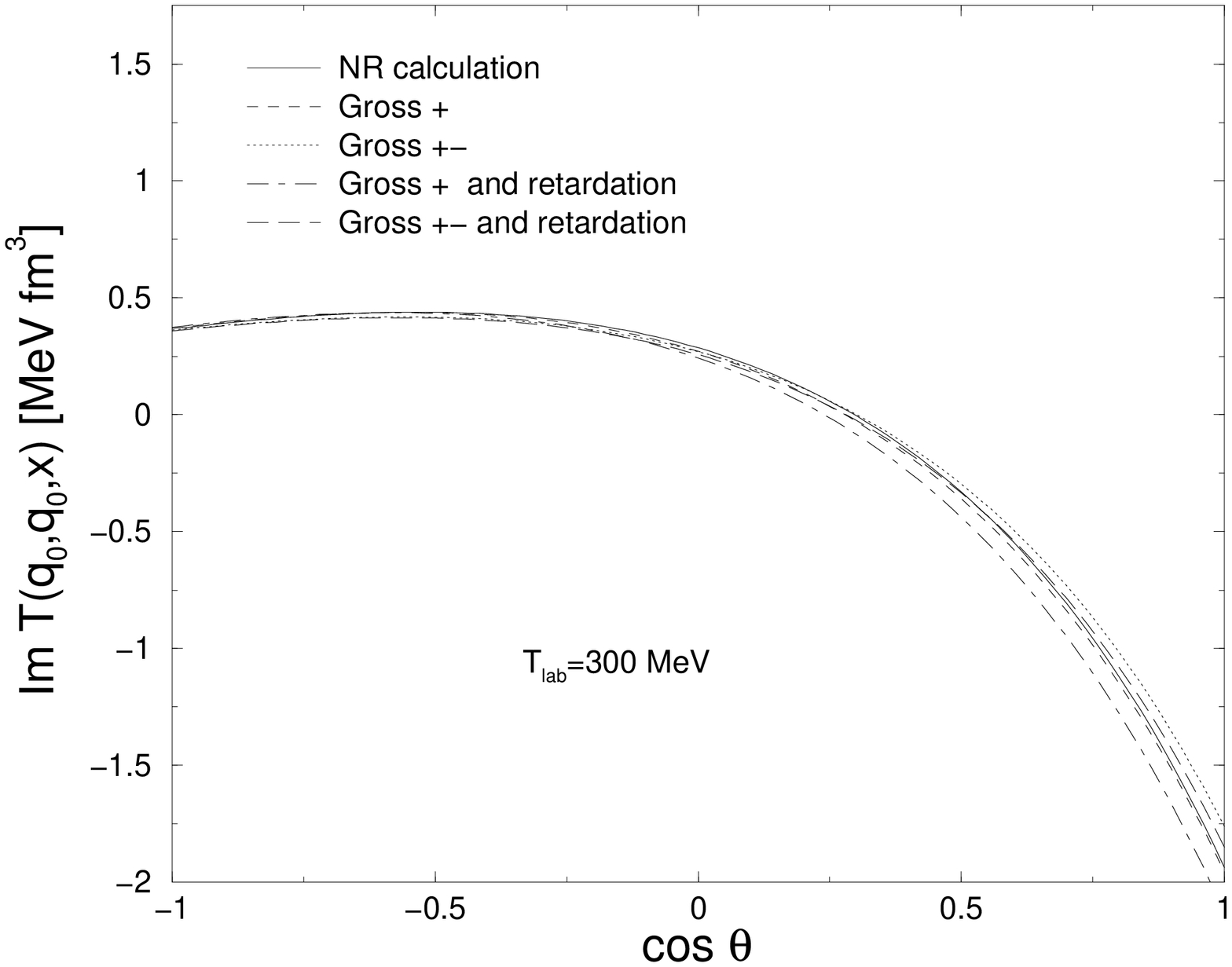, width=6.5cm, angle=270}} 
\centerline{
\epsfig{file=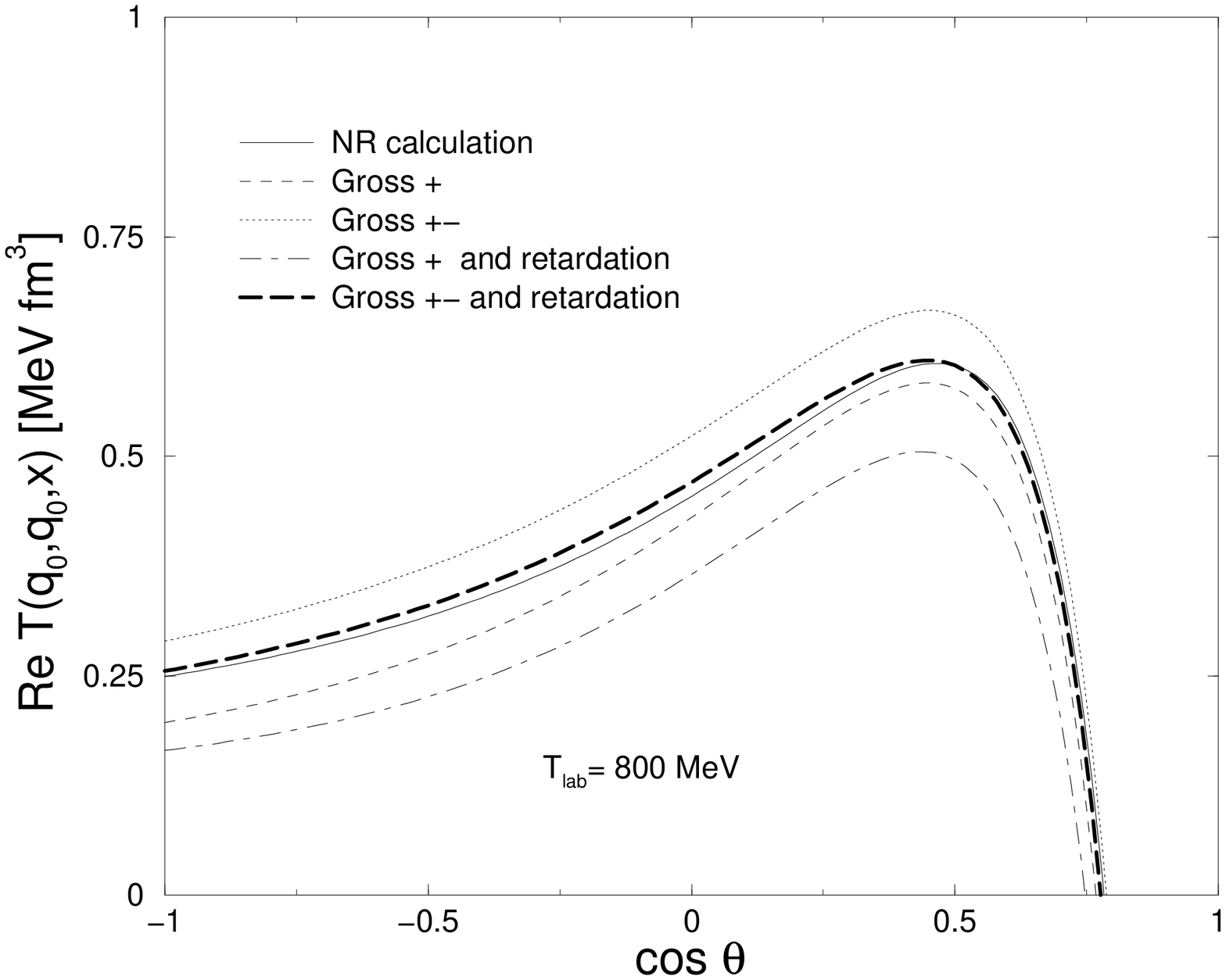, width=6.5cm, angle=270} 
\epsfig{file=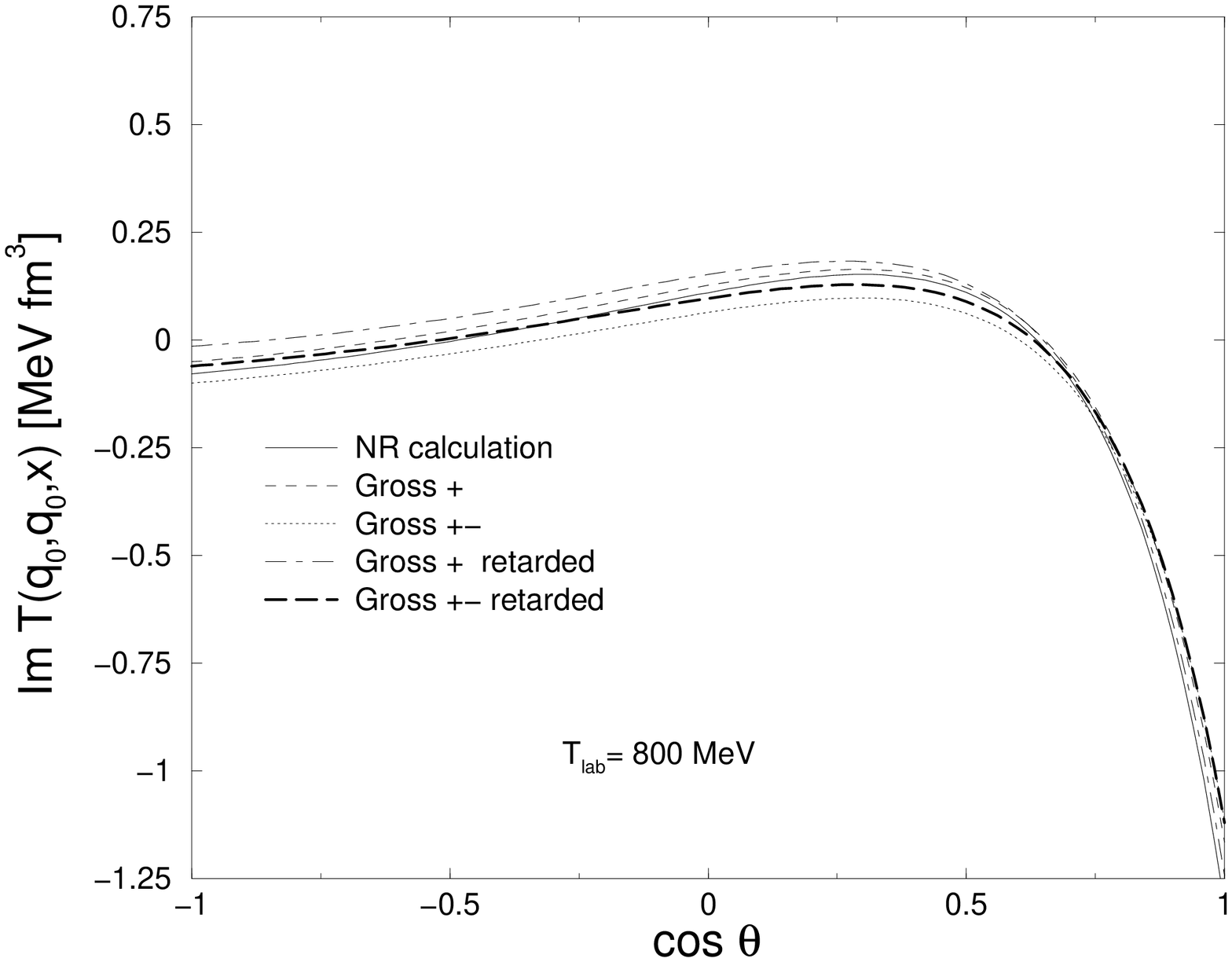, width=6.5cm, angle=270}} 
\vspace{-0.5cm}
\caption{300 MeV and 800 MeV scattering amplitude. 
Retardation and negative energy state effects.}
\end{figure}
%---------------------------------------------------------------------

To understand the previous results we analysed 
the relativistic effects included in the spectator/Gross equation, one by one. 
We decomposed the nucleon propagator in that equation into its positive
and negative energy component components
\begin{equation}
g_{Gross}(k;W)=i 2\pi \frac{1}{2}
\left(\frac{M}{E_k}\right)^2 
\left[ \frac{1}{2E_k-W-i\varepsilon}+
\frac{1}{W} \right] \delta(k_0 +W/2-E_k).
\end{equation}

The results of the decomposition are presented in Fig.\ 3. 
The negative energy effects (Gross $+-$) are essential 
and increase with energy. On the other 
hand, retardation contributions (Gross + and retardation) 
have the opposite effect. Most importantly,
one concludes also that there is an almost perfect cancellation
between retardation and negative energy state effects,
with the net result that the spectator amplitude calculated with retardation  
and the NR amplitude are almost coincident. 
This agreement is not perfect and deteriorates with energy.

Ref. \cite{Elster00} has extended in that direction the NR work of Ref.\
\cite{Elster98},  by considering a realistic NN interaction.
We have in progress a covariant two-nucleon 
calculation which deals with the complete nucleon structure 
arising from its nature as a Dirac particle.

\end{document}